\documentclass[10pt,letterpaper,twocolumn]{article} 

\usepackage{times,mathptmx}
\usepackage{amsmath,amssymb,epsfig,picinpar,graphics,float,amssymb,verbatim}
\usepackage{ol2}
\usepackage{color}

\begin{document}

\twocolumn[

\title{Stable dark and bright soliton Kerr combs can coexist in normal dispersion resonators}
\author{Pedro Parra-Rivas$^{1,2}$, Dami\`a Gomila$^{2}$, and Lendert Gelens$^{1,3,*}$}
\affiliation{$^1$Applied Physics Research Group, Vrije Universiteit Brussel, 1050 Brussels, Belgium\\
$^2$IFISC institute (CSIC-UIB), Campus Universitat de les Illes Balears, E-07122 Palma de Mallorca, Spain\\
$^3$Laboratory of Dynamics in Biological Systems, KU Leuven, Department of Cellular and Molecular Medicine, University of Leuven, B-3000 Leuven, Belgium\\
$^{*}$ Corresponding author: lendert.gelens@kuleuven.be}





\begin{abstract}
Using the Lugiato-Lefever model, we analyze the effects of third order chromatic dispersion on the existence and stability of dark and bright soliton Kerr frequency combs in the normal dispersion regime. While in the absence of third order dispersion only dark solitons exist over an extended parameter range, we find that third order dispersion allows for stable dark and bright solitons to coexist. Reversibility is broken and the shape of the switching waves connecting the top and bottom homogeneous solutions is modified. Bright solitons come into existence thanks to the generation of oscillations in the switching wave profiles. Finally, oscillatory instabilities of dark solitons are also suppressed in the presence of sufficiently strong third order dispersion.
\end{abstract}
]

\maketitle


\noindent 
The Lugiato-Lefever \cite{lugiato_spatial_1987} equation (LLE) has attracted  a lot of interest in the last few years for describing the generation of Kerr frequency 
combs in high-Q microresonators driven by a continuous-wave (CW) laser \cite{delhaye_optical_2007,kippenberg_microresonator-based_2011}. These frequency combs can be integrated on chip \cite{brasch_2016} and used
to measure time intervals and light frequencies with a exquisite accuracy, leading to numerous key applications \cite{okawachi_octave-spanning_2011,papp_microresonator_2014,ferdous_spectral_2011,herr_universal_2012,pfeifle_coherent_2014}.
In this framework a Kerr frequency comb corresponds to the frequency spectrum of a temporal dissipative structure, such as patterns or solitons,
circulating inside the cavity \cite{coen_modeling_2013,chembo_spatiotemporal_2013}. While most theoretical studies have focused on the anomalous second-order 
group velocity dispersion (GVD) regime \cite{Leo_OE_2013,Parra-Rivas_PRA_KFCs,godey_stability_2014}, where the typical dissipative states are bright solitons, the normal GVD regime has recently attracted interest due to the difficulty of obtaining anomalous GVD in some spectral ranges. In contrast to the anomalous regime, dark solitons are found in the normal GVD regime, i.e. low-intensity dips embedded in a high-intensity
homogeneous background. The bifurcation structure and temporal dynamics of these dark solitons, also called "platicons", have been recently studied \cite{Lobanov_1,Parra-Rivas_dark1,Parra-Rivas_dark_long}, and their 
origin is intimately related with the locking of switching waves (SWs) connecting co-existing homogeneous state solutions of high and low intensity \cite{Parra-Rivas_dark1,Parra-Rivas_dark_long}. The generation of such dark pulse Kerr frequency combs has been achieved experimentally by several groups \cite{liang_generation_2014,huang_mode-locked_2015,xue_mode-locked_2015}.

In the anomalous GVD regime it was shown that high-order chromatic dispersion effects can modify the dynamics and bifurcation structure of solitons in the LLE \cite{GelensOL2010}. In particular, third-order dispersion (TOD) generates the 
emission of dispersive waves that can lead to the suppression of dynamical regimes such as oscillations and chaos \cite{milian_soliton_2014,Parra-Rivas_TOD}. As TOD also breaks reversibility, solitons move with a constant velocity\cite{mussot_optical_2008, Leo_prl_2013, jang_observation_2013, Parra-Rivas_TOD, milian_soliton_2014}. While recent work has numerically shown that TOD induces similar dispersive waves in dark solitons \cite{TODnormal_Wang}, studies of the influence of TOD in the normal GVD regime remain scarce. 

In this letter we present a detailed analysis of the bifurcation structure of soliton Kerr combs in the normal dispersion regime in the presence of TOD. In particular we show that stable bright and dark solitons coexist over an increasingly wide parameter range for increasing values of TOD, and we explain such soliton stabilization by analyzing the shape of SWs of different polarities whose locking lies at the heart of this phenomenon. Using the normalization of \cite{leo_temporal_2010}, the LLE reads
\begin{equation}\label{LLE}
 \partial_t u=-(1+i\theta)u+iu|u|^2+u_0-i\partial^2_{\tau}u+d_3\partial^3_{\tau}
\end{equation}
where $t$ is the slow time describing the evolution of the intracavity field $u(t,\tau)$ on the time scale of the cavity photon lifetime and $\tau$ is the fast time
that describes the temporal structure for the field on the time scale of the resonator round trip $L$.
The first term on the right-hand side describes cavity loses(the system is dissipative by nature); $u_0$ is the amplitude of the homogeneous (CW) driving field
field or pump; $\theta$ measures the the cavity frequency detuning between the frequency of the input pump and the nearest cavity resonance; $\partial^2_{\tau}$
models the GVD (here assumed to be normal at the pump frequency); $d_3\partial^{3}_{\tau}$ models the TOD and $d_3$ is its strength; and the sign of the cubic term is set so that it corresponds
to the self-focusing Kerr nonlinearity.

\begin{figure}
\centering
\includegraphics[scale=1]{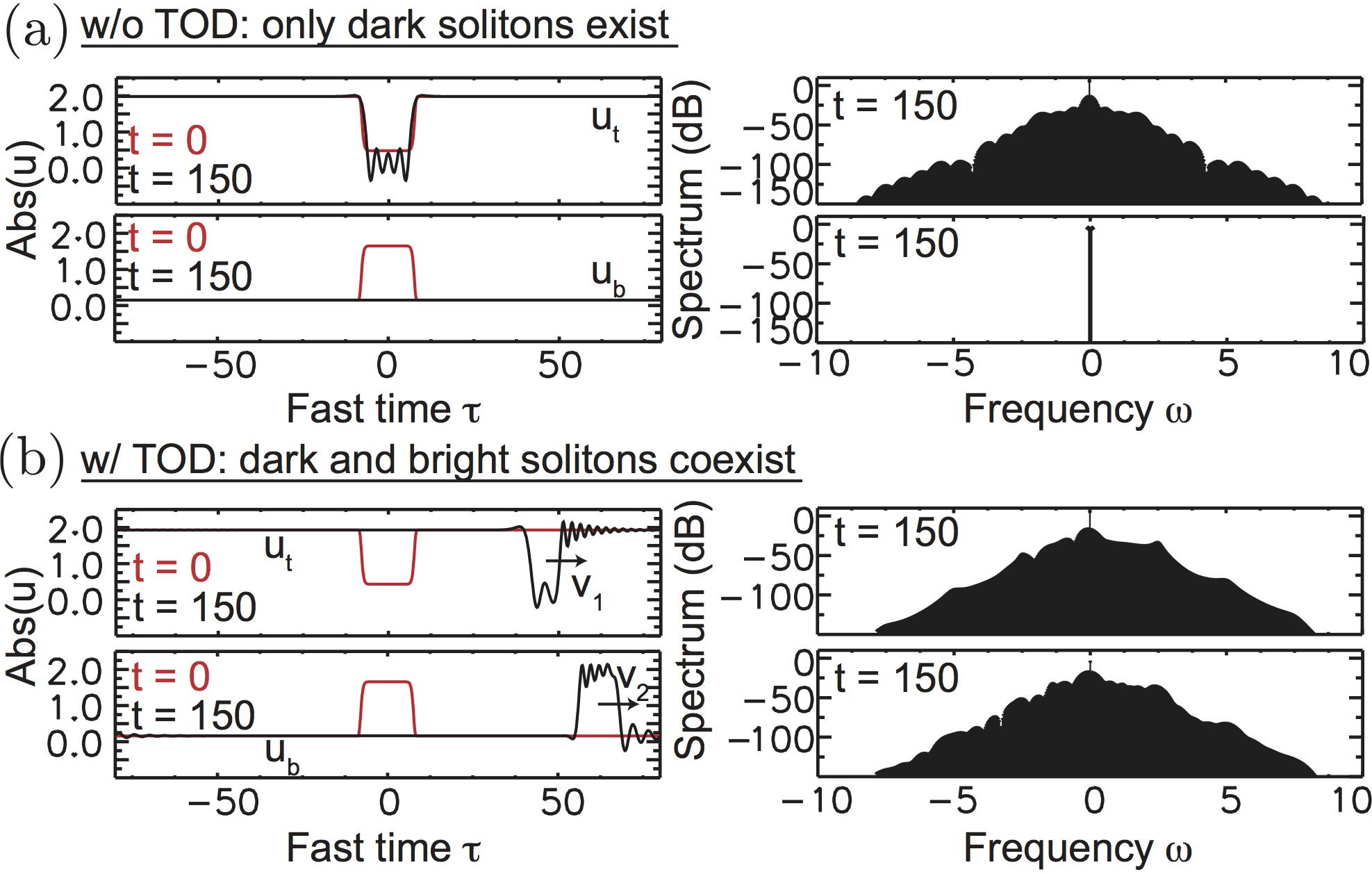}
\caption{Solution profile at $t=150$ (black) after time evolution in the LLE (\ref{LLE}) of an initial
dip in the top HSS $u_t$ or a bump on the bottom HSS $u_b$ (red) in the absence (a) or presence of TOD (b). 
Left panels show the time profile, while the right panels show its associated comb spectrum. 
Parameter set:  (a) $(\theta,u_0,d_3)=(4,2.175,0)$; (b) $(\theta,u_0,d_3)=(4,2.3,0.7)$ .}
\label{Figure1}
\end{figure}

For $\theta>\sqrt{3}$, three homogeneous steady state (HSS) solutions exist: $u_t$ (a stable HSS of higher intensity), $u_m$ (an unstable saddle HSS of intermediate intensity), and $u_b$ (a stable HSS of lower intensity). Figure \ref{Figure1} illustrates how the system can relax into different stable structures depending on the initial condition. In the absence of TOD, Figure \ref{Figure1}(a) shows that a dip in the high intensity HSS $u_t$ (red) can evolve into a stable dark soliton (black), while a bump on the low intensity HSS $u_b$ (red) relaxes back to $u_b$. This observation that dark solitons exist, but bright solitons do not, is general \cite{Lobanov_1,Parra-Rivas_dark1,Parra-Rivas_dark_long}. Ref.\ \cite{Parra-Rivas_dark1} discussed that dark solitons exist due to the locking of overlapping oscillatory tails in the profile of SWs connecting the upper state $u_t$ to the bottom state $u_b$. As such oscillations are absent in SWs approaching the upper state $u_t$, bright solitons do not exist (apart 
from at one single value of the pump, called the Maxwell point $u_0^M$ \cite{Parra-Rivas_dark1}). Figure \ref{Figure1}(b) shows a similar numerical simulation, but now in the presence of TOD ($d_3=0.7$). The initial condition corresponding to a dip still evolves to a dark soliton, which now has an asymmetric profile that moves at velocity $v_1$ while stably maintaining its temporal shape and corresponding frequency spectrum. However, in contrast to the case without TOD, an initial bump now no longer relaxes to the HSS $u_b$, but it forms a bright soliton corresponding to a fixed profile moving at velocity $v_2$).

\begin{figure}
  \centering
  \includegraphics[scale=1.0]{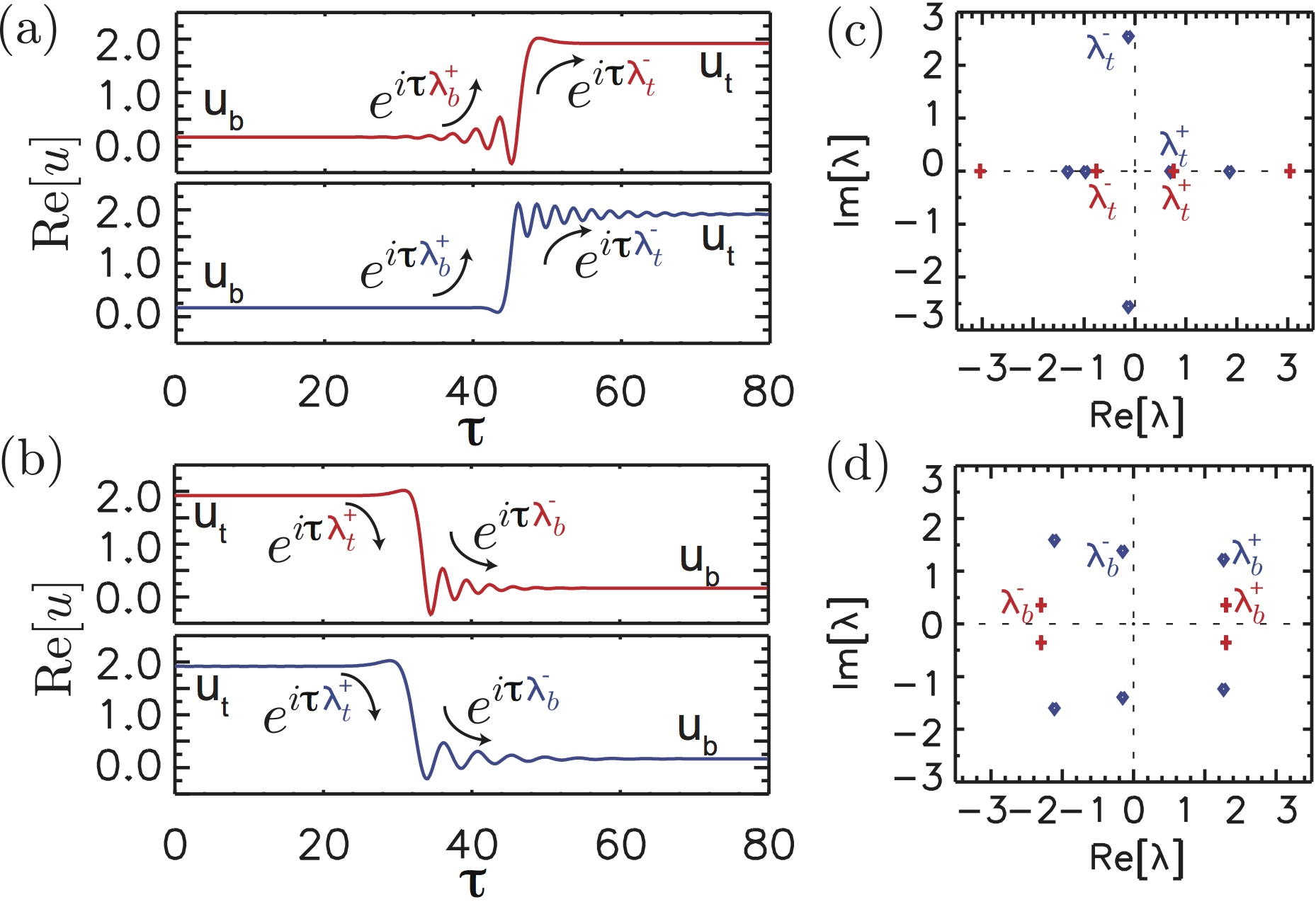}
  \caption{Switching waves $SW_u$ (a) and $SW_d$ (b), in each case for $d_3=0$ (top panel, red) and $d_3=0.7$ (bottom panel, blue). The spatial eigenvalues corresponding to $u_t$ and $u_b$ are shown in (c) and (d), respectively, both  for $d_3=0$ (red crosses) and $d_3=0.7$ (blue diamonds).  $u_0=2.32529$.}
  \label{spatial}
\end{figure}

As already suggested, analyzing the shape of the SWs connecting HSSs $u_t$ and $u_b$ is key to understanding why dark and bright solitons can stably coexist in the presence of TOD. Figure \ref{spatial}(a) shows the shape of an up-switching wave $SW_u$ connecting the low intensity HSS $u_b$ to the high intensity HSS $u_t$, both in the absence (red) and presence (blue) of TOD. Figure \ref{spatial}(b) similarly shows the profile of the opposite down-switching wave $SW_d$, connecting the high intensity HSS $u_t$ to the low intensity HSS $u_b$. Figures \ref{spatial}(c)-(d) plot the corresponding "spatial eigenvalues $\lambda$" of the linearized stationary problem around each HSS, which can be determined analytically \cite{colet_formation_2014,gelens_formation_2014}. The damping rate $q$ and the frequency $\Omega$ of the oscillatory tails around $u_t$ and $u_b$ correspond to the real and imaginary part of certain spatial eigenvalues $\lambda$. The approach to the top HSS $u_t$ by SW$_u$ and the departure from $u_
t$ by SW$_d$ can be approximated linearly in the form $SW_{u,d}(\tau) - u_{t}\propto e^{\lambda_{t}^{-,+} \tau}$, where the $\lambda_t$ is each time the leading spatial eigenvalue. Similarly, the approach to and departure from the bottom HSS $u_b$ is approximated by $SW_{u,d}(\tau) - u_{b}\propto e^{\lambda_{b}^{+,-} \tau}$.

Without TOD, the system is reversible under the transformation $\tau\rightarrow-\tau$. As a consequence spatial eigenvalues (red) come in pairs and are symmetric respect to both axes Re[$\lambda$]=0 and Im[$\lambda$]=0. This means that both SW$_u$ (a) and SW$_d$ (b) approach and leave $u_t$ and $u_b$ in the same way. The top HSS $u_t$ is always approached/left in a monotonic way, explained by the corresponding purely real spatial eigenvalues. In contrast, the bottom HSS $u_b$ is always approached/left in an oscillatory fashion, because its corresponding spatial eigenvalues are complex. Such oscillations can interlock around $u_b$ to form stable dark solitons, but the absence of similar oscillations around $u_t$ prevent stable bright solitons to form. When adding TOD, the dynamics around the HSSs are described by six eigenvalues instead of four, and they are no longer symmetric. The shape of the $SWs$ changes in such a way that they now approach/leave the HSSs $u_t$ in an oscillatory way. Therefore locking 
can occur not only in the bottom HSS $u_b$ but also in the top one $u_t$, forming bright solitons.

\begin{figure}
  \centering
  \includegraphics[scale=1.0]{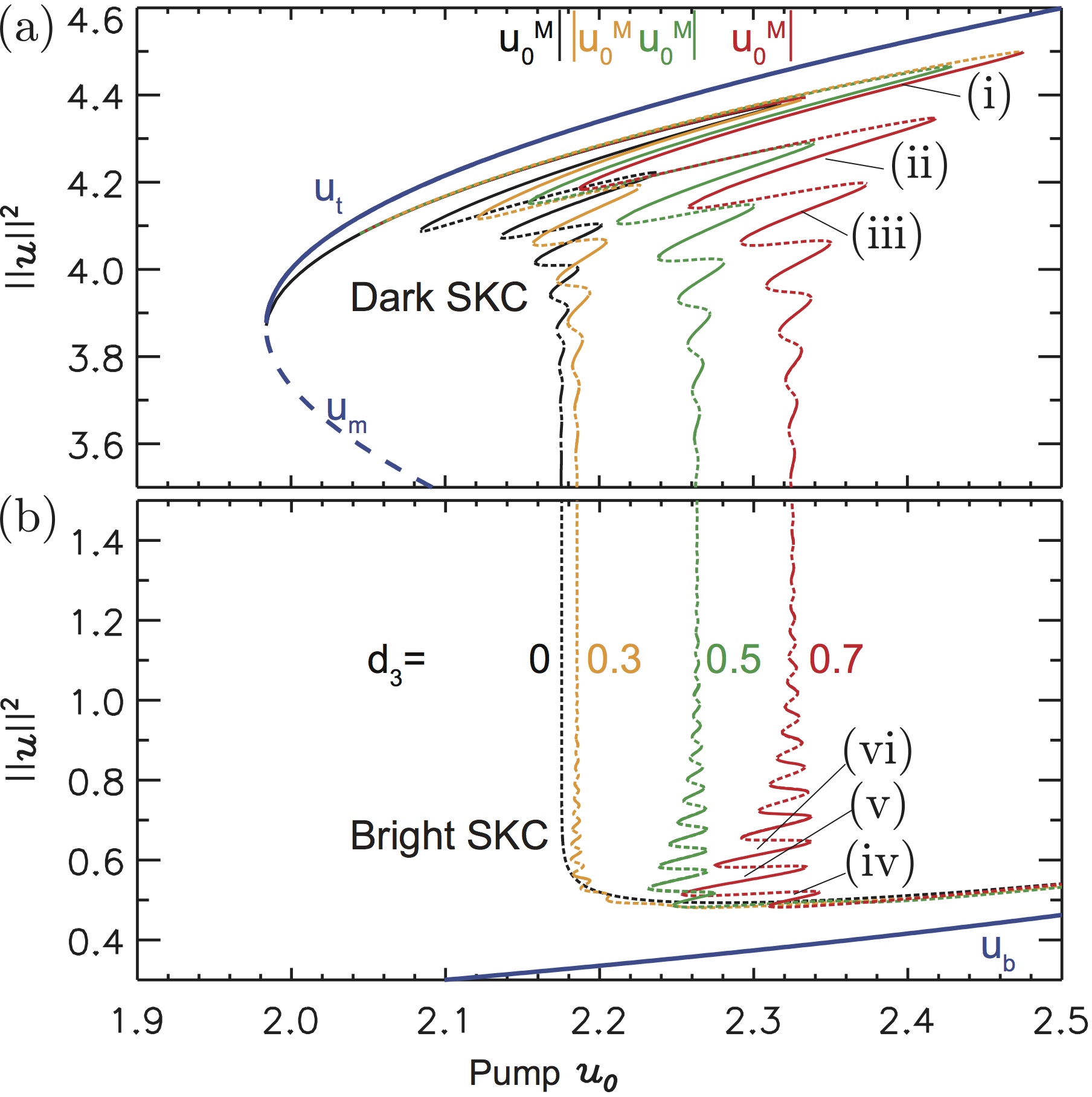}
  \caption{Bifurcation diagrams of HSSs (blue) and dark and bright solitons for $\theta=4$ and increasing values of $d_3 $= 0 (black), 0.3 (yellow), 0.5 (green), 0.7 (red). Solid (dashed) lines correspond to stable
  (unstable) states.}
  \label{diagrams}
\end{figure}

\begin{figure}
  \centering
  \includegraphics[scale=1.0]{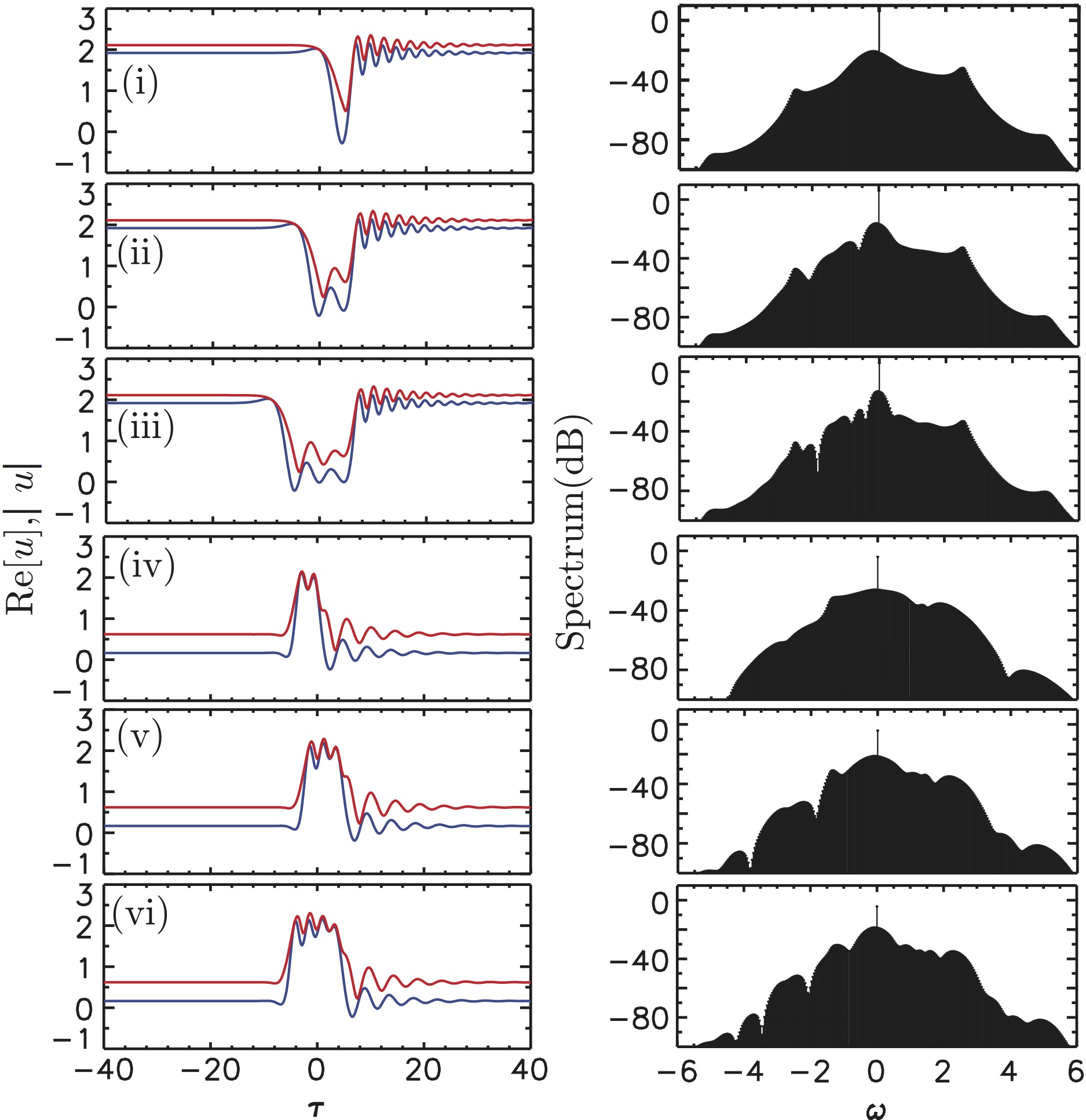}
  \caption{Temporal profiles [left; Re($u$) in blue, $|u|$ in red] and spectral intensities (right, in dB) of asymmetric 
  dark and bright solitons corresponding to the locations (i)-(iii) and (iv)-(vi) in Fig.~\ref{diagrams}. $(\theta,u_0,d_3)=(4,2.232,0.7)$.}
  \label{dark}
\end{figure}

In Fig.~\ref{diagrams} we plot, using the mean energy $||u||^2\equiv L^{-1}\int_0^L|u|^2d\tau$, the bifurcation diagrams of HSSs and solitons in function of the pump amplitude $u_0$ for $d_3=0$ and increasing values of $d_3$. In blue, the three HSS solutions are shown. In black, the bifurcation diagram of dark solitons is shown for $d_3=0$ and has been discussed in detail in \cite{Parra-Rivas_dark1,Parra-Rivas_dark_long}. Unstable (dashed line) dark solitons originate from the saddle-node point SN$_{hom,2}$ and acquire stability (solid line) at the next turning point when increasing the pump amplitude. Dark solitons of increasing width, corresponding to branches with lower mean energy, exist over a parameter range ($u_0$) that becomes narrower and narrower, eventually collapsing to the Maxwell point of the system at $u_0^M$. These dark solitons are connected by unstable solution branches that serve to add additional spatial oscillations in their profiles, leading to the broadening of the dark states. This 
type of bifurcation structure is called a {\it collapsed snaking} structure. When TOD is taken into account, $SW_u$ gradually develops oscillations, which allows bright solitons to come into existence. This can be seen in Fig.~\ref{diagrams}(b) in yellow  ($d_3$=0.3) where bright solitons now exist over a narrow range of pump values $u_0$. Increasing $d_3$ further, the spatial oscillations in $SW_u$ become stronger, leading to the stabilization of bright solitons over a wider range of pump parameters ($d_3$= 0.5 (green), 0.7 (red)). Fig.~\ref{diagrams}(a) shows that also the dark solitons exist over a wider range of parameters, and the Maxwell point $u_0^M$ (around which the collapsed snaking structure is organized) shifts to higher values of the pump $u_0$. Typical solution profiles of solitons corresponding to different branches are plotted in Figure \ref{dark} for $d_3$= 0.7 (red). In the presence of TOD, both dark [Figure \ref{dark}(i)-(iii)] and bright [Figure \ref{dark}(iv)-(vi)] solitons exist over an 
increasingly narrow parameter range as they increase their width.

\begin{figure}
  \centering
  \includegraphics[scale=1.0]{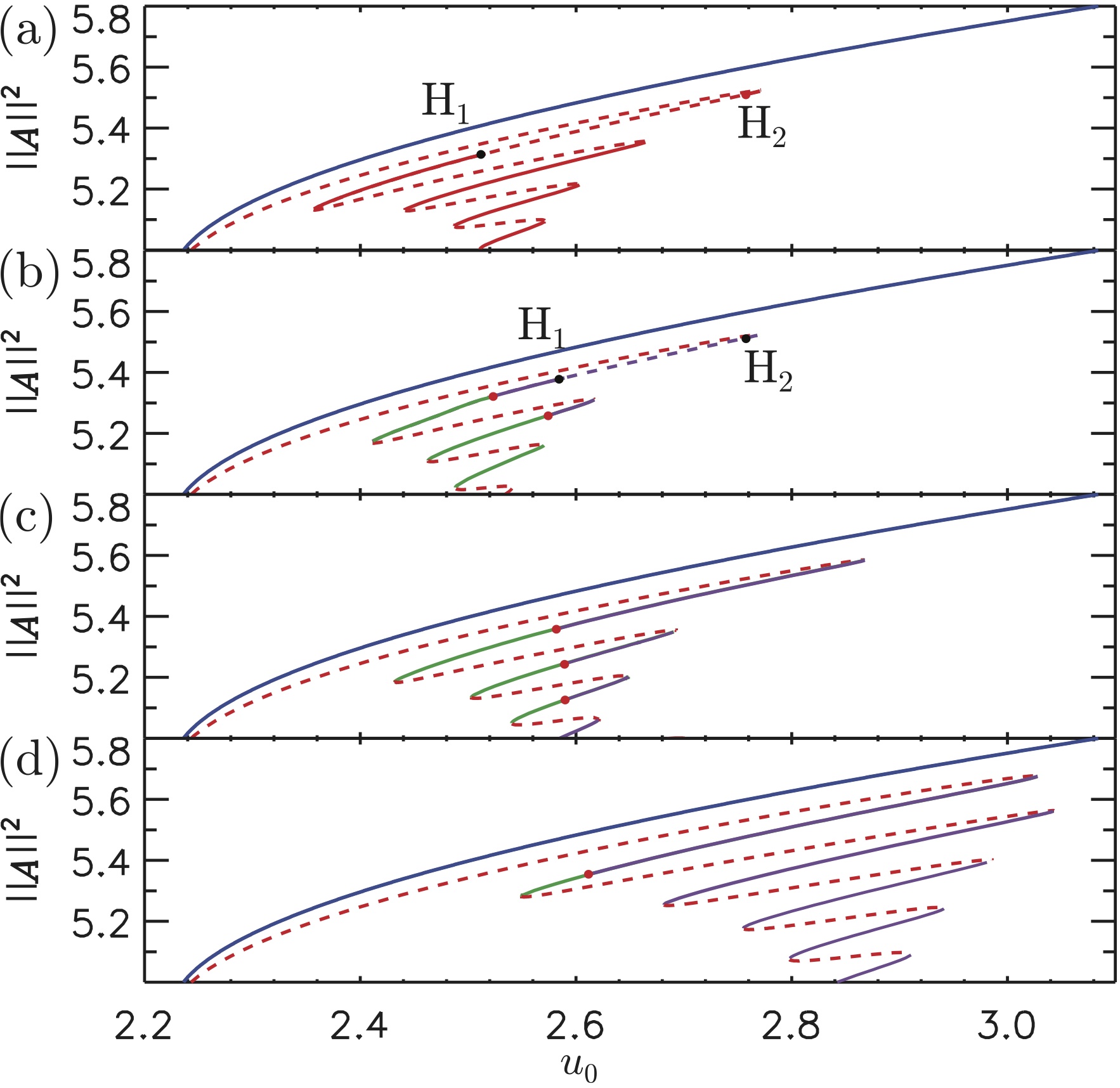}
  \caption{Bifurcation diagrams for $\theta=5$ at different values of $d_3 $= 0 (a), 0.2 (b), 0.3 (c), 0.7 (d). H$_{1,2}$ correspond to Hopf
  bifurcations (black dots). Solid (dashed) lines stand for stable (unstable) states. In solid purple (green) lines we refer to states with positive (negative) velocity. Solid red lines (for $d_3$ = 0) and red dots indicate zero velocity.}
  \label{dark_t5}
\end{figure}

For $d_3 = 0$, at higher values of the detuning $\theta$ dark solitons have been shown to undergo Hopf instabilities and period doubling bifurcations starting a route to temporal chaos \cite{Parra-Rivas_dark1,Parra-Rivas_dark_long}.
This scenario is similar to the one regarding bright solitons in the anomalous dispersion regime \cite{Leo_OE_2013,Parra-Rivas_PRA_KFCs}. In that regime, it was shown that 
TOD, which leads to drift instabilities, could suppress such oscillatory and chaotic temporal dynamics of bright solitons \cite{Parra-Rivas_TOD}. 
We found that this mechanism of stabilization is also present in the normal dispersion regime. 
To illustrate this we first show the bifurcation diagram for $\theta=5$ and $d_3=0$ in Fig.~\ref{dark_t5}(a). Dark solitons 
are unstable between H$_1$ and H$_2$ leading to temporal oscillations. When $d_3\neq0$, the Hopf bifurcations shift in such way that the oscillatory region shrinks for increasing values of TOD until it disappears (see Fig.~\ref{dark_t5}(b)-(d) for $d_3=0.2, 0.3$ and $0.7$ respectively). While for $d_3=0.2$, dark solitons between H$_1$ and H$_2$ oscillate and drift, for $d_3=0.3$ and $0.7$ the oscillatory instabilities have been suppressed and only a drifting soliton remains. The direction in which the soliton drifts is not obvious by looking at its profile and changes with the pump $u_0$. In Fig.~\ref{dark_t5}, solid purple (green) lines correspond to dark solitons with positive (negative) velocity, while the solid red lines (for $d_3 = 0$) indicate zero velocity. By increasing the strength of TOD, the parameter range of solitons with negative velocity shrinks. Although in principle bright solitons could also undergo oscillatory instabilities and similar stabilization with increasing TOD, for the parameter 
range considered in this work, no such oscillations of bright solitons have been found.

\begin{figure}
  \centering
  \includegraphics[scale=1.0]{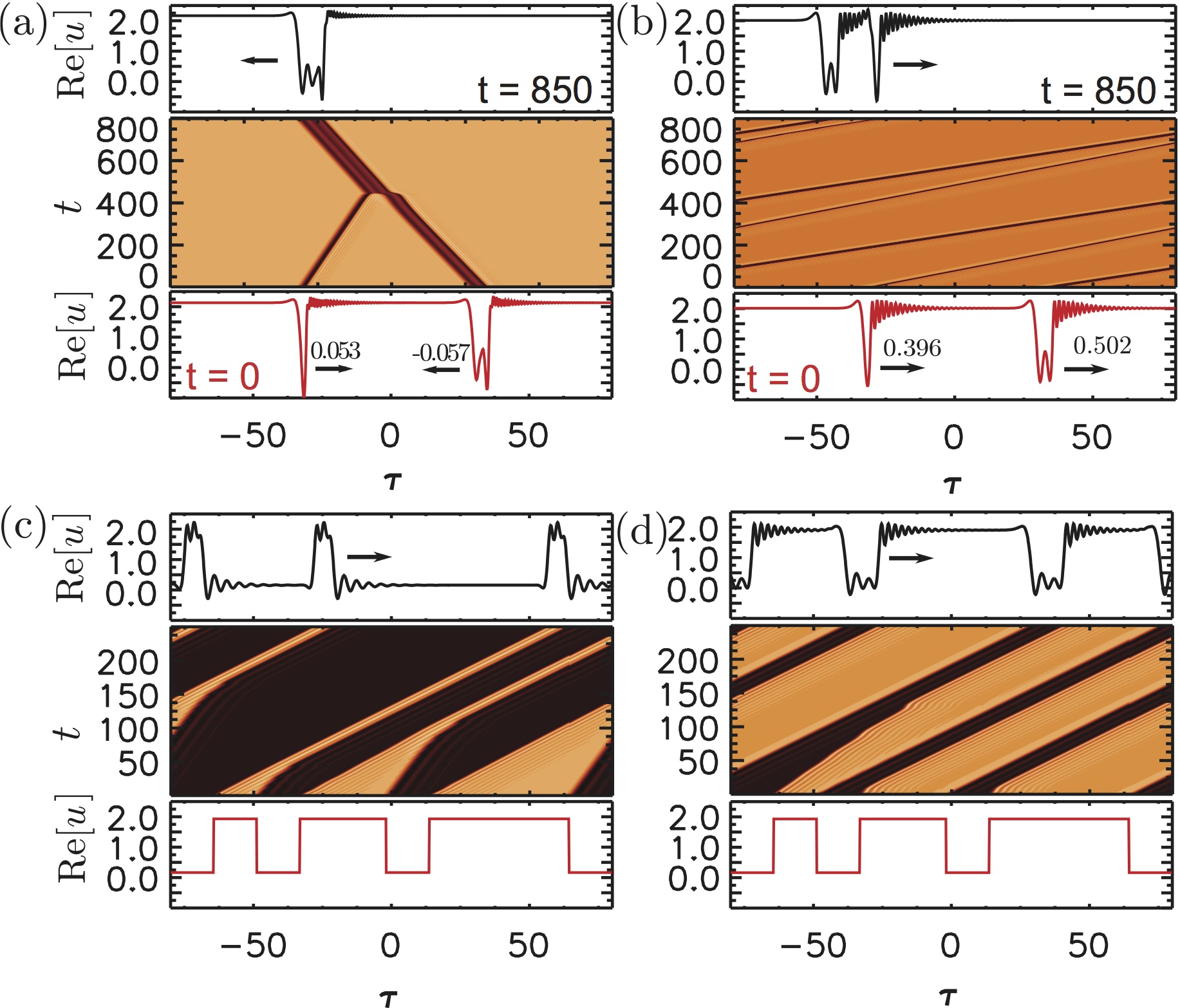}
  \caption{Time evolution in the LLE (\ref{LLE}) of an initial condition (in red) corresponding to: (a)-(b) two dark solitons traveling at different speeds; and (c)-(d) a distribution of three SW$_u$ and SW$_d$ such that the system is in the top and bottom HSS in 50\% of the domain. The profile of the final stable, traveling structure is shown in black. Parameter sets:  (a) $(\theta,u_0,d_3)=(5,2.55,0.2)$,  (b) $(\theta,u_0,d_3)=(5,2.8,0.4)$,  (c) $(\theta,u_0,d_3)=(4,2.26,0.7)$, (d) $(\theta,u_0,d_3)=(4,2.36,0.7)$. }
  \label{interaction}
\end{figure}

The velocity of solitons is not solely determined by the system parameters, but also varies with the width and shape of the soliton. This is illustrated in Fig.\ \ref{interaction}(a)-(b), where two dark solitons of different width travel at different speeds and thus unavoidably interact upon collision. Two dark solitons are dissipative structures and do not come out of such a collision unchanged such as the classical solitons in conservative systems. Instead, a new wider soliton [Fig.\ \ref{interaction}(a)] or a bound state of both solitons [Fig.\ \ref{interaction}(b)] can be created. Fig.\ \ref{interaction}(c)-(d) shows the interaction of several bright and dark domains for two different values of the pump power. In Fig.\ \ref{interaction}(c) the pump is below the Maxwell point $u_0^M$, such that the bottom HSS is favored and a combination of bright solitons result, while in Fig.\ \ref{interaction}(d) the pump is above $u_0^M$, such that the top HSS is dominant and the formation of dark solitons is favored.

%
%


In summary, we have presented a bifurcation analysis of solitons and their corresponding Kerr combs in the normal GVD regime in the presence of TOD. Bright solitons have been shown to be stable over a wide parameter region due to
locking of SWs. TOD can create such oscillatory tails in the SWs around the high intensity HSS $u_t$. Both dark and bright solitons are organized in a collapsed snaking bifurcation diagram, such that broader solitons always exist over a narrower parameter region. Furthermore, TOD can suppress oscillatory and chaotic instabilities of dark solitons in a similar fashion than for the anomalous case. Finally, we have shown that solitons of different widths propagate at different velocity, and multiple different solitons in a resonator thus eventually collide inelastically to form new solitons.

We acknowledge support from the Research Foundation--Flanders (FWO-Vlaanderen) (PPR), the Belgian Science Policy Office (BelSPO) under Grant IAP 7-35, the Research Council of the Vrije
Universiteit Brussel, the Spanish MINECO and FEDER under Grant ESOTECOS (FIS2015-63628-C2-1-R) (DG).


\end{document}